\documentclass[aps,prl]{revtex4}  
\usepackage{bm}
\usepackage{graphicx}
\usepackage{amsmath}
\usepackage{amssymb}
\usepackage[dvips]{color}
\usepackage{subfigure}
\begin{document}
\title{Optimal percentage of inhibitory synapses in multi-task learning} 
\author{Vittorio Capano$^{1}$, Hans J. Herrmann$^{2,3}$ and Lucilla de Arcangelis$^{4}$}
\affiliation{
$^{1}$ Physics Department, University of Naples
Federico II, Napoli, Italy\\
$^{2}$  Institute Computational Physics for Engineering Materials,
ETH, Z\"urich, CH\\
$^{3}$ Departamento de F\'{\i}sica, Universidade Federal do Cear\'a,
60451-970 Fortaleza, Cear\'a, Brazil\\
$^{4}$ Department of Industrial and Information Engineering,
Second University of Naples and INFN Gr. Coll. Salerno, Aversa (CE), Italy
Corresponding author: lucilla.dearcangelis $@$ unina2.it}

\begin{abstract}
{ 
Performing more tasks in parallel is a typical feature of complex brains.
These are characterized by the coexistence of excitatory and inhibitory synapses, whose
percentage in mammals is measured to have a typical value of 20-30\%.
Here we investigate parallel learning of more Boolean rules in neuronal networks. We find that
multi-task learning results from the alternation of learning and forgetting of the individual rules. 
Interestingly, a fraction of 30\%  inhibitory synapses optimizes the overall performance,
carving a complex backbone supporting information transmission with a minimal shortest path length. 
We show that 30\% inhibitory synapses is the percentage maximizing the learning performance
since it guarantees, at the same time, the network excitability necessary to express the response
and the variability required to confine the employment of resources.}
\end{abstract}


\maketitle

Brain functions, such as learning and memory, operate through coordinated neuronal activations
in which highly connected neurons (hubs) have an orchestrating role. Synaptic plasticity \cite{1}
regulates the balance of excitation and inhibition shaping cortical networks into
a complex scale-free functional structure, where hubs are found to be inhibitory neurons  \cite{2}.
Experimental observations suggest that in mammalian brains the fraction of
inhibitory synapses is close to 20-30\%, however this value is not justified by any theoretical
argument.
Evoked activity in the mammalian cortex exhibits a large variability to a repeated stimulus, 
which is expression of the fluctuations in ongoing activity when the stimulus is applied  \cite{3}.
Spontaneous activity therefore represents the fundamental support on which neuronal systems
develop learning skills. In recent years, a novel mode of spontaneous activity has been detected, 
neuronal avalanches. These are bursts of firing neurons measured both in vitro and in vivo, whose
size and duration distributions show a robust power law behaviour \cite{4,5,6}.  
In this framework, 
learning can be interpreted as a phenomenon occurring in a critical state, where the ongoing
activity does not exhibit a characteristic size. Recently, learning of single Boolean rules has 
been investigated \cite{7} in a neuronal network model able to fully reproduce the scaling 
properties and the temporal organization of neuronal avalanches \cite{8,9,10}. 
The model, without any reinforcement learning \cite{11} or error back propagation \cite{12}, 
is able to learn even complex rules: The learning performance increases monotonically with
the number of times a rule is applied and all rules can be learned (and remembered) provided 
that the plastic adaptation is sufficiently slow. The performance rate and the learning time
exhibit universal scaling features, independent of the particular rule and, surprisingly, 
the percentage of inhibitory  synapses in the network. 
Multi-task learning requires a more complex organization of neuronal activity where inhibitory
synapses may indeed play an important role since competing rules have to
establish a synaptic backbone for information transmission under interference effects, not
present in single rule learning. Synaptic plasticity is the key process sculpting this backbone.
Recently, interesting homeostatic features have been detected in plasticity consisting in balanced
depression and potentiation of excitatory and inhibitory synapses aiming at the conservation of 
the total synaptic strength \cite{1}.


Here we study the parallel learning of Boolean rules with two
inputs (AND, OR, XOR and RAN, a rule associating to each input configuration a random output)
on a neuronal network
able to reproduce the statistical behaviour of spontaneous activity.
A firing (non-firing) neuron corresponds to the Boolean variable 1 (0) (see Methods). The
response of the network to the successive applications of each rule is monitored for
several trials, where plastic adaptation is performed according to a non-uniform negative feedback
algorithm if the system gives a wrong answer (see Methods). We demonstrate that the
30\% fraction of inhibitory synapses, measured in mammalian brains, 
optimizes the network's performance. We provide an understanding
of this behaviour in terms of the structural features of the cluster of paths
supporting information transmission and the activity dynamical properties.


In the following we will investigate the learning performance of the model as function of the
parameter  $p_{in}$. The role of other parameters has been addressed in previous works
\cite{27,28} and found to
be irrelevant for the scaling behaviour of avalanche activity.
We first verify that for each rule
the choice of hubs inhibitory neurons and homeostatic plastic adaptation optimize the performance 
in both single and parallel learning (see Suppl. Fig.1,2). On this basis we adopt these two
ingredients in our procedure. Moreover, we verify that also for parallel learning the performance
improves for increasing average connectivity in the network  and for decreasing distance
between input and output neurons. Most importantly, slow plastic adaptation improves the performance 
also in multi-task learning and curves corresponding to different plastic responses 
collapse onto a universal function by appropriately rescaling the axes in terms of the plastic 
adaptation strength (Suppl. Fig.3).


\section{Results}
We first analyse the evolution of the performance in learning more rules in parallel. 
Surprisingly, this exhibits a percentage of success for each rule that
does not increase monotonically with the number of trials, as for single rule learning \cite{7}
(Fig.1). In the case of learning in parallel the OR and AND rules the non-monotonicity is dramatic: 
Already after few applications of both rules in parallel the network recovers the right answer  
to OR (blue circles)
with a performance of about 100\%, however it cannot provide the right answer to AND (red triangles)
and is therefore unable to learn both rules (black line). After about 500 trials, the network starts
recovering the right answer to AND, however this leads to a sharp decrease in the 100\% OR performance, 
down to about 50\% performance for both rules,  corresponding to a random 0/1 answer.
After further applications, the network improves again its performance for each individual rule, which
implies that it has to partially forget the right answer to OR in 
order to learn the AND rule. A good performance
of 90\% in learning both rules is reached only after a longer training time. The above observation 
suggests that some rules (OR in the previous case) are "easier" to learn and will drive the
initial performance. However the network needs a longer training time to organize the synaptic
connections able to provide the right answer to both rules.
This effect is also observed for three-rule-learning, where oscillations are less
striking but well beyond statistical fluctuations. In this case the easiest rule to learn is the random rule
(non-monotonicity  shown in the inset),
whereas AND and XOR learning exhibit similar performances. In both two and three-rule learning, 
the performance in the whole task is always monotonic. This suggests that the final response in 
complex learning
results from the alternation of learning and forgetting, where partially "unlearning" is functional
to the improvement in the overall performance.

We then question the dependence of the multi-task learning performance on the 
percentage of 
inhibitory synapses $p_{in}$ in the network. Surprisingly, the success rate does not show a 
monotonic trend with $p_{in}$ (Fig.2). In particular the asymptotic value of the performance
strongly depends on $p_{in}$ and exhibits a maximum
for $p_{in}\simeq 20-30\%$. Increasing the network size improves the asymptotic performance, 
as in single rule learning \cite{7}, but does not affect the non-monotonic behaviour.
Interestingly, the initial configuration of synaptic strengths does not affect the performance results.
In particular, even starting from a uniform distribution of strengths for all neurons, at the end
of the learning routine inhibitory neurons are on average stronger than excitatory ones and, for
$p_{in}\simeq 30\%$, the average strength ratio is close to typical balanced networks \cite{26}.
Therefore, the plastic adaptation regulates the relative strength and hinders the creation of 
excessively strong inhibitory synapses. 

To better understand this result, we first analyse the structure of the union
of  all paths of active synapses connecting input and output neurons which provide the right answer
to all rules. This constitutes the backbone supporting information  transmission.
Each backbone in Fig.3 shows the temporal sequence of firing neurons, where neurons active simultaneously
lie on the same horizontal line and the temporal evolution describes how firings propagate in time
from input neurons (green) at $t=0$ to the output neuron (black). 
The lines connect the pre-synaptic firing neuron with the 
post-synaptic stimulated neuron drawn at its own firing time. Therefore, a one-time-step line indicates
a strong temporal correlation between successive firings, whereas long temporal connections evidence
a temporal delay between the pre and post-synaptic neuron firings. The last time step contains always the
output neuron, that can be also reached at earlier times, namely more than once during an avalanche. 
Fig.3 shows that for a purely excitatory network
and for  $p_{in}=50\%$, the backbone has a simpler structure involving few neurons, whose 
firings are mostly successive in time. Conversely,
for $p_{in}=30\%$ the structure becomes larger and with an intricate network of firings.
Indeed the size of the backbone averaged over many configurations exhibits a maximum value
for this fraction of inhibitory synapses (Suppl. Fig.4). Moreover, the average
shortest path connecting input and output neurons within the backbone exhibits
an opposite trend for different $p_{in}$, namely  a minimum value for
$p_{in}\simeq 20-30\%$ (Suppl. Fig.4). 
Furthermore, for these values of $p_{in}$ the multiplicity of independent paths through each 
neuron (Suppl. Fig.5) varies over a wider range, indicating
the emergence of alternative paths, enhancing cooperative effects under the combined
action of excitatory and inhibitory synapses. At the same time, within this complex
backbone a preferential path for information transmission is identified under the combined
action of excitatory and inhibitory synapses.
The fraction $p_{in}\simeq 30\%$ thus optimizes the structural features.

Recent experimental results have suggested that neuronal systems in a resting  state can be
viewed as systems  acting close to a critical state where activity does not have a characteristic
size. We question if the same behaviour is observed for the
response to an external stimulus. We measure the  size of neuronal avalanches as the number of neurons
active when the system provides the right answer to each step of all rules.  
For purely excitatory networks learning is achieved only if all
neurons are involved in the avalanche, i.e., the distribution is peaked at $s*=N$ (vertical line Fig.4). 
However, this does not imply that all  neurons belong
to the backbone, because many neurons are activated in an inefficient attempt to reach the
output neuron. Conversely, for increasing $p_{in}$, activity is
progressively confined, namely the peak height decreases and the distribution extends towards smaller
sizes. This implies that the system can recover the right answer by involving a
limited number of neurons (Fig.4) and therefore by a more efficient employment of resources.
Results suggest that purely excitatory networks, able to exhibit avalanches without a characteristic size 
in spontaneous activity, cannot provide the right answer to a stimulus unless the entire system is involved.
This observation is particularly striking in the case of a right answer "zero". 
In the following, we will provide an understanding of these optimal learning conditions.

An avalanche of size $s^*$ represents a configuration able to learn the rule where $s^*$ neurons are
active and $N-s^*$ are inactive. In the framework of spin models, this avalanche
is therefore analogous to a configuration with $s^*$ up spins and $N-s^*$ down spins. 
The probability $P(s^*)$ to obtain an $s^*$-avalanche, regardless of the individual firing neurons,
is the probability to observe such configuration.
Here $p_{in}$ tunes the degree of structural disorder in the system and
therefore plays the role of temperature in thermal systems.  
We then define the entropy associated to the learning dynamics
 as 
$$S =-\sum_{s^*} P(s^*)\ln P(s^*)\eqno(1)$$ 
which quantifies the variability in the
response provided by the system. The entropy $S$ is equal to zero for $p_{in}=0\%$, as for
thermal systems at zero temperature, and progressively 
increases with $p_{in}$ (Fig.5).
In a learning routine the network must be able to trigger and sustain the activity in
response to the external stimulus, which is, for instance, impossible is a purely inhibitory network. 
As a consequence, the level of excitability of the whole
network must be adequate to tackle the task. We quantify the level of excitability of the network
by evaluating the average synaptic strength $E=<g_{ij}>$, where $g_{ij}$ is positive 
(negative) for excitatory (inhibitory) synapses. The excitability $E$ is a decreasing function
of $p_{in}$ and tends to zero for $p_{in}\to 50\%$ (Fig5).
In order to combine the above ingredients, we propose a novel functional 
$$F=E p_{in}S\eqno(2)$$ 
which quantifies the balance between variability, both in the structure and in the response, and the excitability of the system.
This definition is reminiscent of the free energy in statistical physics, 
with the additional requirement that the system cannot learn in a state with zero excitability. 
As the free energy, $F$ is composed of an energetic and an entropic term and vanishes 
in the extreme cases of 
absence of variability or excitability, assuring that both features are required for learning.
This functional is a non-monotonic function of $p_{in}$ and
exhibits a maximum  for $p_{in}\simeq 30\%$ (Fig.5), percentage that leads to optimal learning by
the ability of the system to react to an external stimulus combined to the possibility of tuning the 
response saving resources. 



\section{Discussion}

Real brains are able to perform more tasks in parallel. We investigate the multiple task
learning 
performance of a neuronal network able to reproduce the critical behaviour of spontaneous activity.
Networks undergo the teaching procedure of two or more rules in parallel, via a numerical procedure
(different inputs and the same output for all rules) which allows, at the same time, to introduce
neuronal firing interference and to monitor separately the performance to each rule. 
The network starts by learning the "easiest" rule, the one which requires
less plastic adaptation. Under this point of view, OR is the easiest rule to learn, whereas XOR
requires a longer training. Even if the first rule is learned, further applications stimulate the 
system to accomplish the entire task. The network is then obliged to redefine the connectivity network, 
even if it was successful in providing the right answer to the first rule. 
This operation implies that the system
partially forgets the previous right answer, down to a level of performance where 
the answer is compatible to a random outcome. Only after this partial forgetting the system is
able to identify a synaptic structure providing the right answer to all rules. Therefore,
learning and forgetting appear to be the two, apparently opposite, mechanisms that must coexist to
realize multi-task-learning.

Current evidence from functional magnetic resonance imaging (fMRI) experiments \cite{13,14}
and EEG data \cite{15,16} shows that a greater brain signal variability indicates a more
sophisticated neural system, able to explore multiple functional states. Signal variability also 
reflects a greater coherence between regions and a natural balance between excitation and 
inhibition originating the inherently variable response in neural functions. 
Furthermore, the observation that older adults exhibit less variability reflecting less 
network complexity and integration, suggests that variability can be a proxy for optimal systems. 
In our study, we quantify variability in the response by the Shannon entropy associated to neuronal
activity in learning, combined to the structural disorder measured by the 
percentage of inhibitory synapses in the system. The quantity $p_{in}S$ is therefore an entropic 
term measuring the level of variability, far from being just noise. 
At the same time, the cognitive
performance must rely on the capability of the system to react to an external stimulation.
This feature can be interpreted as an energetic term, which is
maximum for purely excitatory networks, where learning always triggers an 
all-encompassing activity, involving also unnecessary resources. 
Therefore, it is beneficial for brains performing complex tasks to realize a balance between 
the entropic and the energetic features.

Our study shows that  multi-task learning is optimized not for a structurally balanced network
with 50\% inhibitory synapses, but for $30\%$ inhibitory synapses, value measured
in mammalian brains. This percentage allows the desired balance between excitability and 
variability in the response. To better understand why this particular value complies with the two 
main requirements for learning, excitability and variability, it is useful to recall the structure
of the backbone of paths carved by the plastic adaptation process. Indeed, the backbone results
in a more complex structure of firing connections for $30\%$ inhibitory synapses, than for a structurally
balanced network: The number of neurons involved usefully in the process increases. Indeed, a purely 
excitatory network involves all neurons in successfully performing a task but the majority of them
does not operate to convey the information from input to output, i.e. the backbone is quite bare.
At the same time the multiplicity of paths going through each neuron varies over a range which is
maximum for $p_{in}=30\%$. This does not simply imply longer paths, 
since the average length of the shortest path connecting input and output is minimum for $p_{in}=30\%$.
Also this result is striking since it would be more reasonable to expect shorter
paths for a higher percentage of inhibitory synapses, corresponding to a multiplicity range an order
of magnitude smaller.
Results suggest that a complex backbone, with a wide range of path multiplicity and therefore
the coexistence of a larger number of paths, including very short ones, is the optimal
firing structure for learning. This observation confirms that learning is a truly collective process,
where a large number of units with intricate connections participate to the dynamics.
An excess of excitatory or inhibitory synapses, with respect to this optimal value, would hinder the
emergence of this complex structure and therefore limit the learning performance of the system. 
It would be interesting to extend the present study to other networks, as for instance modular
networks, and  verify, as found for spontaneous activity, if the particular network topology
affects the multi-task learning behaviour.

\section{Methods}

{\bf Neuronal network.}
We consider $N$ neurons placed at random in a two-dimensional
space and characterized by a potential $v_i$.
Each neuron can be either excitatory or inhibitory, according to Dale's law,
with a random out-going connectivity degree, $k_{out}$, assigned
according to the experimental distribution of functionality networks \cite{17},
$n(k_{out})\propto k_{out}^{-2}$ with $k_{out}\in[2,100]$.
This distribution implies that the network does not exhibit a characteristic
connectivity degree. On the contrary, few neurons are highly connected and act as hubs
with respect to information transmission, whereas the majority of neurons are connected
to few other neurons.
Connections are established according to a distance dependent probability,
$p(r)\propto e^{-r/r_0}$, where $r$ is their spatial distance and $r_0\simeq16$
the connectivity spatial range \cite{18}.
Once the outgoing connections are chosen, we evaluate the in-degree of each neuron $k_{in}$.
The initial synaptic strengths are randomly drawn from a uniform distribution, 
$g_{ij}\in [0.5,1]$, where $g_{ij}\neq g_{ji}$. Since the neuronal connectivity degree 
is power law distributed, the level of inhibition is expressed in terms of $p_{in}$,
the fraction of inhibitory synapses in the network, with
inhibitory neurons highly connected ($k_{out}>10$) \cite{2}.

A neuron $i$ fires as soon as its
potential is above a fixed threshold,
$v_i \geq v_{\rm max}=6.0$, changing the membrane potential of post-synaptic neurons proportionally
to  $g_{ij}$ \cite{8,9}
$$
v_j(t+1)=v_j(t)\pm \frac{v_i k_{out}^i}{k_{in}^j}\frac{g_{ij}(t)}{\sum_k g_{ik}(t)}\eqno(3)
$$
where $k_{in}^j$ is the in-degree of neuron $j$, the sum is extended to all out-going
connections of $i$ and the plus or minus sign
is for excitatory or inhibitory synapses, respectively.
The factor $k_{out}^i/k_{in}^j$ makes the potential variation of neuron $j$, 
induced by neuron $i$, independent of the connectivity of both neurons, whereas 
the factor $\frac{1}{\sum_k g_{ik}(t)}$ normalizes the synaptic strength values for each neuron.
We note that the unit time step 
represents the time unit for propagation from one neuron to the connected ones, which
in real systems  could be of the order of 10ms.
After firing, a neuron is set to a zero resting potential and
in a refractory state lasting $t_{ref}=1$ time step.
The initial neuron potentials are uniformly distributed between
$v_{\rm max} - 1$ and $v_{\rm max}$. A small random fraction
$10\%$ of neurons is fixed at zero potential and act as boundary sites or sinks for the charge.

{\bf Learning routine.}
For each rule we choose at random two input neurons and a unique output neuron,
under the condition that they are not boundary sites and they are mutually separated on 
the network by $k_d$ nodes. $k_d$ represents the minimum distance between input and 
output neurons, which can also be
joined by much longer paths. In the standard multi-layer perceptron framework, $k_d$ would play the role of
the number of hidden layers. Assigning  the same output neuron to all rules enables,
at the same time, to discriminate between different rules and
to have interference among the different paths carrying information.
We test the ability of the network to learn in parallel
different Boolean rules: AND, OR, XOR and a rule RAN, which associates a random output to all possible 
binary states at two inputs.
For each rule the binary value 1 is identified with the neuron firing, i.e.,
$v_{\rm output}\ge v_{\rm max}$ at any time during the avalanche propagation.
Conversely, the binary state 0 corresponds to the neuron which has been depolarized but
fails to reach the firing threshold during the entire avalanche.

Once  the input sites are stimulated, their activity may bring to threshold other
neurons and therefore lead to an avalanche of firings. We impose no restriction on the number
of firing neurons in the propagation and
let the avalanche evolve to its end.
If the avalanche during the propagation did not reach the
output neuron,
we consider that the system was in a state unable to respond to the given stimulus,
and as a consequence to learn. We
therefore increase uniformly the potential of all neurons by a small quantity,
$\beta = 0.01$, until the configuration reaches a state where the output neuron is first
perturbed.
We then compare the state of the output neuron with the desired output.
A single learning step requires the application of the entire sequence of states for each rule,
letting the activity propagate till the end and then monitoring the state of
the output neuron.
If the answer is right for all three entries of a rule,
this has been learned. The routine proceeds until the system learns all rules.

{\bf Plastic adaptation.}
Plastic adaptation obeys a non-uniform generalization \cite{7}
of the negative feedback algorithm \cite{23}:
If the output neuron is in the correct state according to the rule,
we keep the value of synaptic strengths. Conversely, if the response is wrong, we
modify the strengths of those synapses active during the avalanche \cite{24}
by $\delta g =\pm \alpha/d_k$, where $d_k$ is the chemical distance of each presynaptic neuron
from the output neuron and $\alpha$ represents the strength of synaptic adaptation.
Here $\alpha$ represents the ensemble of
all possible physiological factors influencing synaptic plasticity.
Therefore synapses can be either strengthened or weakened depending on the mistake:
If the output neuron fails to be in a firing state, we strengthen the synapses, 
conversely strengths are weakened if the right answer 0 is missed.
Once the strength becomes smaller than a threshold,
$g_{ij}<10^{-4}$, the synapse is pruned. This ingredient is very important as
since decades the crucial role
of selective weakening and elimination of unneeded connections in adult  learning  has
been recognized \cite{19,20}.

The synapses involved in the signal propagation and responsible for the wrong
answer, are therefore not adapted uniformly but inversely proportional
to the chemical distance from
the output site. Namely, synapses directly connected to the output neuron receive the
strongest adaptation $\pm \alpha$.
This rule dependence on $d_k$ models the feedback to the wrong answer triggered
in the region of the output neuron and propagating backward towards the input neurons.
In the brain this mechanism is regulated by the release of messenger molecules, as some hormones
(dopamine suppressing LTD \cite{21} or adrenal hormones enhancing LTD \cite{22}) or 
nitric oxide \cite{25}.
Three different procedures for plastic adaptations are tested: Homeostatic, uniform and
restricted. In homeostatic plasticity the adaptation of excitatory and inhibitory synapses have
opposite signs to realize conservation of the average strength \cite{1}. In the uniform case all
active synapses undergo the same adaptation, whereas in the restricted case only excitatory 
synapses are modified.

{\bf Acknowledgments.}
We thank the European Research Council
(ERC) Advanced Grant 319968-FlowCCS  for financial support.

{\bf Author contributions}
L.d.A. and H.J.H. were involved in all the phases of this study.
V.C. performed the numerical simulations and prepared the figures.
All authors reviewed the manuscript.

{\bf Competing financial interests:}
The authors declare no competing financial interests.

\medskip
{{\bf Figure 1.  Alternation of learning and forgetting in multi-task performance.}
Percentage of networks learning two (left) or three (right) rules
as a function of the number of times the rules are applied for 500 configurations
with $N=1000$ neurons ($k_d=3$, $\alpha=0.001$, $p_{in}=0.3$).
Left: Performance in parallel learning the OR and AND rules.
Right: Performance in learning three rules in parallel (OR, AND and RAN).
Inset: Performance for the RAN rule at a finer temporal scale.}

\medskip
{{\bf Figure 2. Inhibitory synapses control the non-monotonic performance
in multi-task learning.}
Percentage of networks giving the right answer to both the AND and XOR rules
as a function of the number of times the rule is applied for 500 configurations
with $N=1000$ neurons and different $p_{in}$ ($k_d=3$, $\alpha=0.001$).
Inset: Asymptotic value of the performance vs. $p_{in}$ for
different $N$.}

\medskip
{{\bf Figure 3. Complexity of the structure carrying information depends on the
percentage of inhibitory synapses.} Backbone carrying information
for one configuration with $N=40$ neurons learning the AND and XOR rules with
different $p_{in}$. Neurons
involved in avalanches giving the right answer to both rules are listed
according to their firing time (downwards arrow).
Input (output) neurons are green (black) and
excitatory (inhibitory) neurons are red (blue). Neuron size is proportional to
$k_{in}$. The three-digit number associated to each neuron is composed by the
firing  time followed by the neuron label (two digits).
Long-range links appear if the post-synaptic neuron
fires some time steps after receiving the stimulation from the pre-synaptic neuron. }

\medskip
{{\bf Figure 4. Inhibitory synapses confine the system response in learning.}
Distribution of avalanche sizes  measured when the system gives the right answer to,
both, the AND and
XOR rules for 500 configurations of networks with $N=1000$ neurons and different $p_{in}$
($k_d=3$, $\alpha=0.001$). }

\medskip
{{\bf Figure 5.  Percentage of inhibitory synapses observed in mammalian brains
optimizes the combination of excitability and variability in the response.}
The average synaptic strength and the entropy (left scale) evaluated for 500 
configurations of networks
with $N=1000$ neurons learning the AND and XOR rules are shown as function of $p_{in}$
 ($k_d=3$, $\alpha=0.001$).
The functional $F=E p_{in}S$ (right scale) has a maximum  for $p_{in}=30\%$.   }

\newpage
\vskip 3cm
\begin{figure}
  \rotatebox{0}{\resizebox{.7\textwidth}{!}{\includegraphics{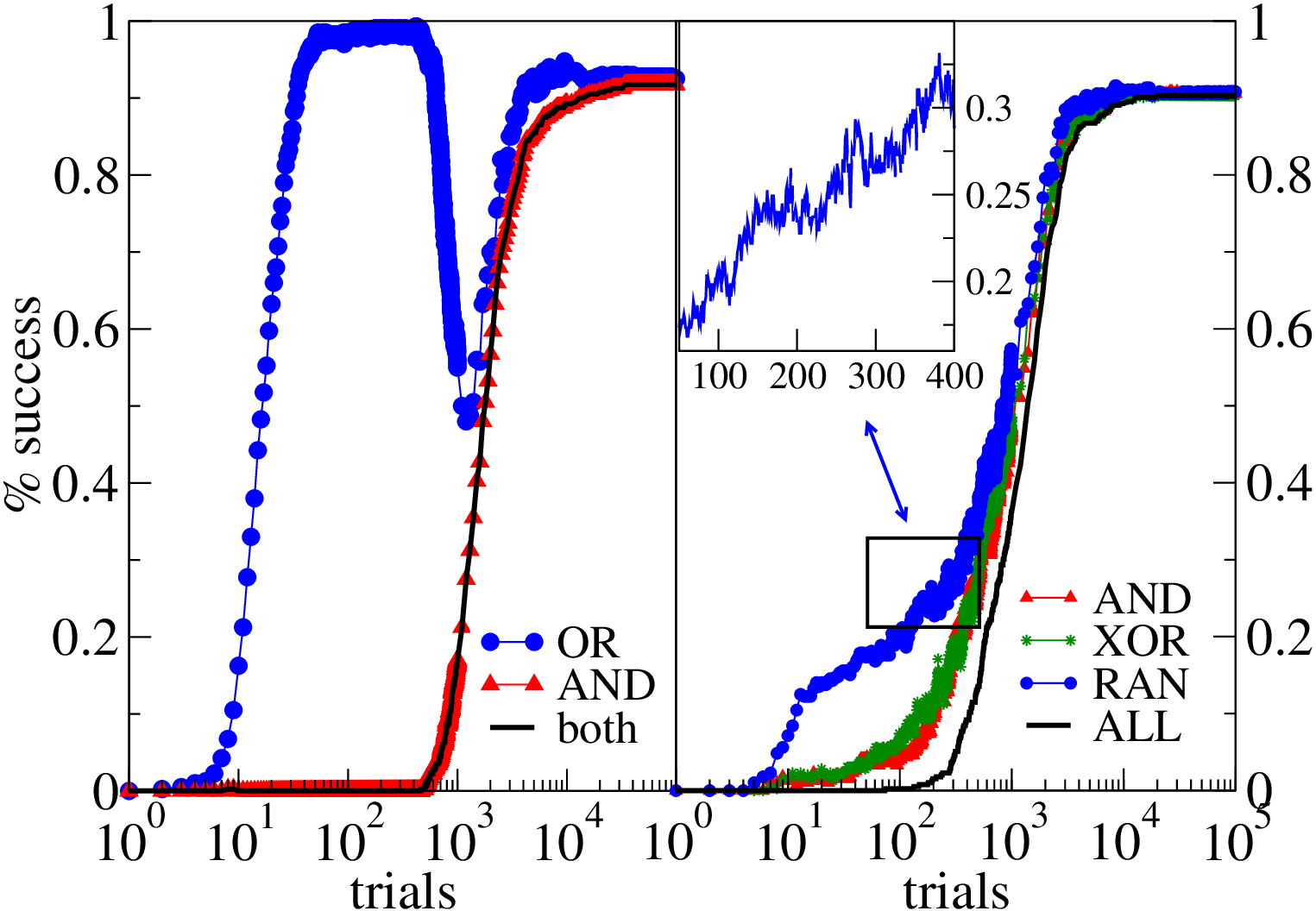}}}
\end{figure}
\vfill

\newpage

\begin{figure}
\vskip 3cm
  \rotatebox{0}{\resizebox{.6\textwidth}{!}{\includegraphics{fig3.eps}}}
\end{figure}
\vfill
\newpage
\begin{figure}
\vskip 3cm
  \rotatebox{0}{\resizebox{.5\textwidth}{!}{\includegraphics{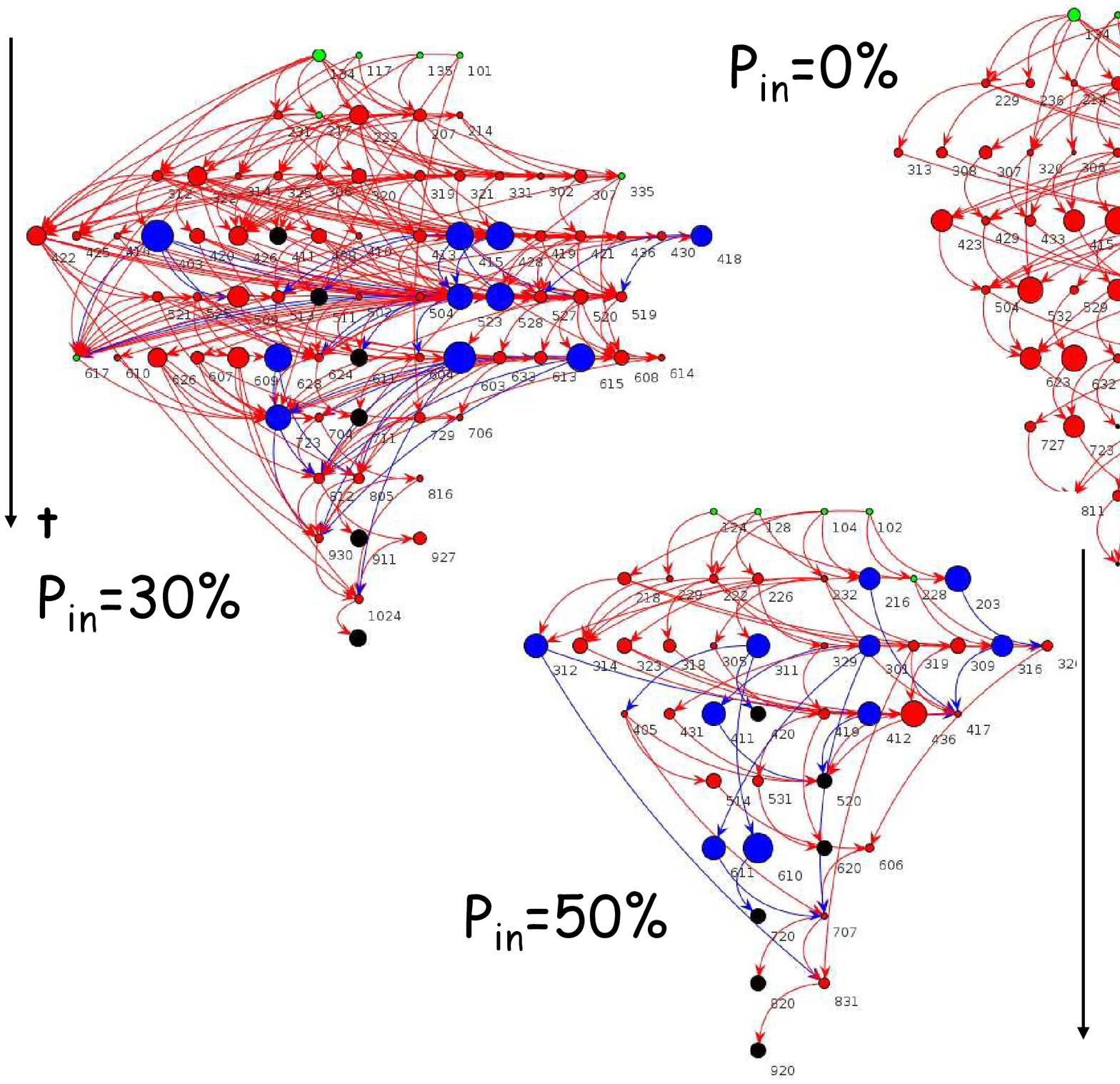}}}
\end{figure}
\vfill
\newpage

\begin{figure}
\vskip 3cm
  \rotatebox{0}{\resizebox{.6\textwidth}{!}{\includegraphics{fig5.eps}}}
\end{figure}

\eject
\newpage

\begin{figure}
\vskip 3cm
  \rotatebox{0}{\resizebox{.6\textwidth}{!}{\includegraphics{fig6.eps}}}
\end{figure}

\end{document}